\documentclass[epj]{svjour}

\usepackage{graphicx}
\usepackage{fancyhdr}
\def\makeheadbox{{
 }}
\def\mytitle{The dark matter as a light gravitino} 
\def\myauthors{Gilbert Moultaka}  
\def\mytype{parallel session}
\def\mysession{Cosmology}

\newcommand{\lsim}{\raisebox{-0.13cm}{~\shortstack{$<$ \\[-0.07cm] $\sim$}}~}
\newcommand{\gsim}{\raisebox{-0.13cm}{~\shortstack{$>$ \\[-0.07cm] $\sim$}}~}
\begin{document}
\title{The dark matter as a light gravitino}
\author{Gilbert Moultaka
\thanks{ based on
work 
in collaboration with
K. Jedamzik ({\sl LPTA-Montpellier}), M. Lemoine ({\sl IAP-Paris}) 
\cite{JLM04}, \cite{JLM05}, and  work in progress, M. Kuroda (Meiji-Gakuin), M. Lemoine (Paris), M. Capdequi-Peyranère
(Montpellier).}%
}                     
%
%
\institute{Laboratoire de Physique Th\'eorique et Astroparticules \\
{\sl UMR5207--CNRS}, Universit\'e Montpellier II \\
Place E. Bataillon, F--34095 Montpellier Cedex 5, France
}
%
\date{}
\abstract{We address the question of gravitino dark matter in the context of gauge 
mediated supersymmetry breaking models. A special emphasis is put on the role
played by the MSSM singlet messenger in the case of $SO(10)$ grand unification. 
} 
\maketitle
\section{Introduction}

In some instances, the requirement that supersymmetric particle dark matter 
scenarios aught to be the simplest to handle cosmologically and the least
model-dependent, seems occasionally to take  over the more fundamental
question  for supersymmetry, namely the origin of supersymmetry breaking. Since
there is  to date no particularly compelling susy breaking mechanism/model to
be  preferred to all the others, one should also consider the dark matter issue
from a particle physics standpoint which offers different classes of susy 
breaking mechanisms, irrespective of whether the ensuing cosmological context 
is ``simple" or not.  

Recent developments  \cite{ISS}, \cite{murayama} stressing the
existence of metastable susy breaking vacua,   have renewed the interest in
gauge-mediated susy breaking (GMSB) scenarios opening new possibilities
for the model-building  \cite{GR99}, and appear to be  very interesting from
the early Universe point of view as well \cite{AK}.  On the other hand, the
gravitational interactions which play  a minor role for susy breaking in GMSB
models remain   physically relevant through the coupling to supergravity, at
least in order  to absorb  the unphysical goldstino component, to adjust the
cosmological  constant to a small value and to avoid a massless R-axion. 
In such scenarios where supersymmetry breaking is triggered by non-perturbative
dynamics of  some (secluded) gauge sector and communicated to the MSSM by a
messenger sector through  perturbative gauge interactions, the susy breaking
scale $\sqrt{F}$ and the mass scale $\Lambda$ of the  secluded gauge sector can
be well below the Planck scale.
Moreover, if these two scales combine to trigger the electroweak 
symmetry breaking yielding $G_F^{-1/2} \sim (\alpha/4 \pi) k {F / \Lambda } $,
where  $G_F$ is Fermi's constant (and $0< k \le 1$ measures the secludedness
of  the secluded  sector), then the gravitino mass  $m_{3/2} \simeq F/(\sqrt{3}
m_{\rm Pl}) \sim \left(4 \pi/ \alpha)(\Lambda /\sqrt{3} k m_{\rm Pl}\right) 
G_F^{-1/2}$ where $m_{\rm Pl}$  is the reduced Planck mass, is expected to be 
very small ($\lsim {\cal O}(1)$ GeV) and is the lightest supersymmetric
particle (LSP). The question then arises as to which particle can be a good
candidate  for the cold dark matter (CDM) in this case? To answer this question
requires an unconventional treatment as compared to the Neutralino ``vanilla''
candidate or even to the heavy gravitino candidate in the context of gravity
mediated susy breaking models. Indeed, in contrast with the latter where the
hidden sector  is typically too heavy to be produced at the end of inflation,
the secluded and  messenger sectors of GMSB provide stable particles that may
be present in the early Universe for a sufficiently heavy reheat temperature
$T_{RH}$. We consider hereafter such configurations assuming that only the
messenger (including the spurion) sector  can be produced and  illustrate its
relevance to the issue of the CDM.

\section{Curing a Messenger Problem}

The mass degeneracy within a supermultiplet of messenger fields is lifted by 
susy breaking leading to a lighter and a heavier scalar messengers with masses
$M_{A \pm} =M_X(1 \pm {k F/M_X^2})^{1/2}$ and a fermionic partner with mass 
$M_X$ (where $F$ and $M_X$ are related to the dynamical scale $\Lambda$).
 Thus ${k F/M_X^2} < 1$. Moreover, one has to require 
 ${k F/M_X} \lsim 10^5 \mbox{GeV}$ to ensure an MSSM spectrum 
 $\lsim {\cal O}(1) \mbox{TeV}$. One then expects typically
 
 \noindent 
 $M_X \gsim 10^5 \mbox{GeV}$. In GMSB models the lightest messenger 
 particle (LMP) with mass $M_{-}$ is stable due to the conservation of a 
 messenger quantum number. If present in the early Universe the messenger 
 particles are thermalized through their gauge interactions with the thermal 
bath. The corresponding LMP relic density is calculable similarly to the case of 
Neutralino LSP albeit an extended particle content and couplings.  
However, it turns out to be typically too large to account 
for the CDM  even in the most favorable case of the 
electrically neutral component of a $\mathbf{5}+\mathbf{\overline{5}}$ 
representation of $SU(5)$ where it is found to scale as 
$\Omega_M h^2  \simeq 10^5 \left({M_{-}/10^3 TeV}\right)^2$ 
with the LMP mass \cite{DGP}. The situation is even worse in the case of
$SO(10)$ where the LMP is an MSSM singlet with a suppressed annihilation 
cross-section leading to a very large relic density. One possible cure to this 
messenger overcloser problem is to allow the LMP to decay to MSSM particles. 
This can be easily achieved by adding renormalizable but messenger number
violating operators to the superpotential, however, such low-scale operators
would tend to ruin the nice FCNC suppression of GMSB models. A more appealing
approach is to insist on the messenger number conservation as a consequence
of a discrete accidental symmetry at low energy and invoke the typical
violation of such a (non gauge) symmetry by gravitational interactions 
\cite{ross} once the GMSB model is coupled to supergravity. The LMP decay
would then occur via Planck mass suppressed operators in the Lagrangian,
which can originate either directly from effective non-renormalizable
operators in the K\"ahler or the superpotential, or indirectly after susy
breaking through (holomorphic) renormalizable operators in the K\"ahler
potential. In the latter case the suppression is controlled by the gravitino
mass. In the case of $SU(5)$, an exhaustive study of these operators was 
carried out in \cite{JLM04} for messengers transforming as 
$\mathbf{5}+\mathbf{\overline{5}}$ or $\mathbf{10}+\mathbf{\overline{10}}$.  
In $SO(10)$  with one set of 
messengers transforming as $\mathbf{16} + \overline{\mathbf{16}}$,
the LMP decay can be  induced by 
non-renormalizable  operators  
in the K\"ahler potential, e.g. $K \supset \mathbf{16}_F 
\overline{\mathbf{16}}_M^\dag \mathbf{10}_H/m_{\rm Pl}$, 
 or in the superpotential, e.g. $W \supset 
\overline{\mathbf{16}}_M {\mathbf{16}}_F {\mathbf{16}}_F \mathbf{10}_H 
\times m_{\rm Pl}^{-1}$, leading respectively to $2$- and $3$-body decays,
where $ \mathbf{16}_M (\overline{\mathbf{16}}_M), \mathbf{16}_F$ and 
$\mathbf{10}_H$ denote respectively the messenger, the standard matter
and the  electroweak Higgs supermultiplets.
We assume a typical decay width

\noindent
$\Gamma_{\rm LMP} = (1/16\pi) f' M_X^3/m_{\rm Pl}^2$ where $f'$ 
parameterizes our ignorance of the couplings  and possible further phase space 
suppression. For couplings of ${\cal O}(1)$, $f'\simeq 1 ( 3 \times 10^{-3})$
for $2$- ($3$-body) decays into essentially massless particles. 
On the one hand, such suppressed decays  would not upset the  FCNC 
suppression, and on the other, will turn out actually to be a blessing 
regarding a solution to the gravitino overproduction in the early universe, 
eventually allowing the gravitino to account for the {\sl cold} dark matter in 
the context of GMSB models.

\section{Gravitino abundance}

\subsection{the cosmological set-up}

In favorable parts of the parameter space, the LMP late decay into MSSM
particles can release enough entropy to dilute the initial gravitino relic
density down to a level which can  account for the CDM in the Universe even for
very high $T_{RH}$, \cite{FY02}, \cite{JLM04}, \cite{JLM05}. For this to work,
though, the LMP should dominate the Universe  energy density before it decays,
and should decay after the gravitino has  freezed-out from the thermal bath.
The necessary conditions $T_d < T_{MD}, T^f_{3/2}$ [where $T_d, T_{MD},
T^f_{3/2}$ denote respectively the LMP decay and messenger matter domination
temperatures, and the gravitino freeze-out temperature]  is then determined by
the particle properties and annihilation cross-section  and decay width of the
LMP, delineating the favorable parts of the parameter  space. We have studied
this scenario in detail for the case of $SU(5)$  \cite{JLM04} and $SO(10)$,
\cite{JLM05}, \cite{CKLM}. Here we concentrate on the latter case with one set of messengers
transforming as   $\mathbf{16} + \overline{\mathbf{16}}$.  The entropy release
$\Delta S \equiv S_{\rm after}/S_{\rm before}$, diluting the initial gravitino
density, is determined by the temperatures before and after LMP decay and can
be approximated to $T_{MD}/T_d$.  $T_{MD}$ is given by the LMP yield  and mass
($T_{MD}\simeq (4/3)M_{-} \times Y_{\rm LMP}$) and $T_d$ is determined by the
LMP width ($\Gamma_{\rm LMP} \simeq H(T_d)$). The LMP yield $Y_{\rm LMP}$ is 
controlled by the LMP annihilation into MSSM particles which enters the
corresponding Boltzmann equation. Since in our case  the 
LMP is an $SU(5)$ singlet \cite{DGP}, \cite{JLM05}, this annihilation proceeds 
via loop effects of virtual messengers ($A_M, \psi_M$)  and spurion ($S$) 
exchange. We  consider here the leading annihilation 
cross-section into $2$ gluons, fig.1, and parameterize its
thermal averaged as $\langle \sigma_{1{\rm loop}}v\rangle \sim f \times (\alpha_s/4\pi)^2
\kappa^4/s$  where $\kappa$ is the spurion-messenger coupling ($W \supset  
\kappa \hat{S} \mathbf{16}_M\overline{\mathbf{16}}_M$), $\alpha_s$ the strong
coupling constant, $\sqrt{s}$ the C.M. energy and $f$ a form factor depending
on the internal masses and couplings. Neglecting the very heavy GUT sector 
contributions which typically decouple, one finds 
\begin{eqnarray} && f = \frac{32}{\pi} |(3/8) \bar{g}^2 + M_{A_{-}}^2 C_{-} + 
( (3/4) \bar{g}^2 - 1) 
M_{A_{+}}^2 C_{+}  \nonumber \\ &&    + \; 2 M_X^2(  M_{A_{-}}^2 C_{-} + 
M_{A_{+}}^2 C_{+}  + (  s  - 4 M_X^2) C_X -1 ) \; D[s]|^2 \nonumber  
\end{eqnarray}

\noindent  where $D[s]$ denotes the spurion propagator 
$(s - M_S^2 + i \Gamma_S M_S)^{-1}$, $C_{\pm, X}$ are $C_0$
functions scaling as $s^{-1}$,  and $\bar{g}^2 \equiv 4 \pi \alpha_s/\kappa^2$.
Since the messengers in the  loops carry color charges, a substantial mass
splitting occurs between the contributing $A_{-}$ states     and the LMP due to
RGE running from the GUT scale down to $Q^2=M_{-}^2$, as well as from genuine
loop corrections \cite{DGP96}, \cite{JLM05},  leading typically to $M_{A_{-}}
\simeq 3 M_{-}$. Such effects, as well as the  running of $\alpha_s$  should be
taken into account for a precise determination of $Y_{LMP}$. On the other hand,
the contribution of the scalar spurion depends on its mass and width. The
spurion can be either  heavier or lighter than the messenger sector. Here we
consider only the former case, where the spurion decays at tree-level into
pairs of messengers or at one-loop into MSSM particles.\footnote{If the spurion
is lighter than the LMP, an efficient tree-level annihilation of the latter
into a pair  of spurions would lead to a too small $Y_{LMP}$.} We find that the
decay into MSSM particles dominates irrespective of the value of the coupling
$\kappa$, the total width $\Gamma_S$  remaining small though ($\Gamma_S/M_S
\simeq (1 - 4) \times 10^{-3}$). A careful treatment (beyond the relative
velocity expansion) of the thermally averaged  annihilation cross-section  is
thus required close to this narrow resonance, i.e. typically when $M_{-} \simeq
M_S/2$ and assuming a non-relativistic decoupling of  the LMP from the thermal
bath.

\subsection{relic gravitinos}
When the necessary temperature conditions discussed in the previous section
are met,  the final 
gravitino relic density is given by $\Omega_{{}_{grav}} = \Omega_{{}_{grav}}^{th}/\Delta S + 
\Omega_{{}_{grav}}^{{}^{Mess}} + \Omega_{{}_{grav}}^{{}^{NLSP}}$ where the last
two contributions denote non-thermal production through late decays or
scattering. One should also consider various cosmological constraints 
(hotness/warmness, BBN, species dilution, etc...). Let us illustrate the case
first with some effective fixed  values for $f$ and $f'$. 
This is shown in fig.2 taking $T_{RH} \simeq 10^{12}$ GeV, 
see also \cite{JLM04}. 
The  red horizontal shading shows the theoretically excluded region where 
$ k > 1$; the other red shading indicates the region excluded by BBN 
constraints. If the spurion is heavier than the LMP, gravitino cold DM 
(green region) occurs for relatively light
LMPs and $m_{3/2} \sim 1 \mbox{kev} - 10 \mbox{MeV}$. Note that without the
LMP induced entropy dilution, the reheat temperature would have been 
constrained to be several orders of magnitude lower than $10^{12}$ GeV
in order to avoid overcloser for the range of gravitino masses found here.

\noindent
More generally, in 
the models of ref. \cite{GMSB} one finds \cite{JLM05}

\noindent
$\Omega_{grav}h^2 \,\simeq\, 10^3
f^{0.8} \kappa^{3.2} f'^{1/2}\left({M_{-} \over 10^6\,{\rm
GeV}}\right)^{-0.3} \times \left({m_{3/2} \over 1\,{\rm MeV}}\right)$ for 
non-relativistic LMP freeze-out, putting the gravitino relic abundance in the ballpark of WMAP 
results, for $\kappa \sim {\cal O}(10^{-1})$ and typical ranges for $f$ and 
$f'$.  
Considering now the specific one-loop form-factor $f$ given in the previous
section and assuming for the sake of the illustration a spurion much heavier 
than the messenger sector, we still find regions consistent with WMAP results
for $\Omega h^2$ at the 99\% confidence level, e.g. for $M_X = 10^6 - 10^8$
GeV, one has $1.1 \mbox{MeV} < m_{3/2} < 4$ MeV for a three body decay LMP, and
$65 \mbox{keV} < m_{3/2} < 230$  keV, for a two body decay LMP.

For heavier  spurions the annihilation 
into 2 gravitinos through tree-level gravitational interactions 
(see fig.1) becomes significant as its cross-section scales like 
$\langle \sigma v\rangle\,\simeq\, (1/ 24\pi) k^2 M_-^2/(m_{3/2} m_{\rm Pl})^2
$. It can dominate the 1-loop annihilation, eventually saturating the unitarity
limit (the black dashed line in fig.2), thus disfavouring gravitino 
CDM solutions for $M_{-} \gsim 10^8$ GeV.

\section{Conclusion}
Light gravitino can account for  CDM  irrespective of
$T_{RH}$, making it a good DM candidate in GMSB: typically  if $T_{RH} \lsim 10^5 \mbox{GeV}$ then the messengers are 
not produced and thermal gravitinos with $m_{3/2} \lsim 1 \mbox{MeV}$ provide
the right CDM density, while for $T_{RH} \gsim 10^5 \mbox{GeV}$ the messenger
can be present and should be unstable, thus providing a source of entropy 
production that can reduce a thermally overproduced gravitino to a
cosmologically acceptable level.
Moreover, various constraints (e.g. on $T_{RH}$, \cite{SP}, or on 
the gravitino mass \cite{VLHMR}) simply do not apply in the scenarios we 
have illustrated, thus escaping  possible tension with thermal leptogenesis.
Finally, let us mention that such scenarios can potentially allow to 
avoid the recently studied cosmological gravitino problem due to inflaton 
decay \cite{ETY}.    



\begin{figure*}
\label{fig1}
  \includegraphics[width=.8 \textwidth, height=.8 \textheight, keepaspectratio,
bb= -15 460 580 660, clip]{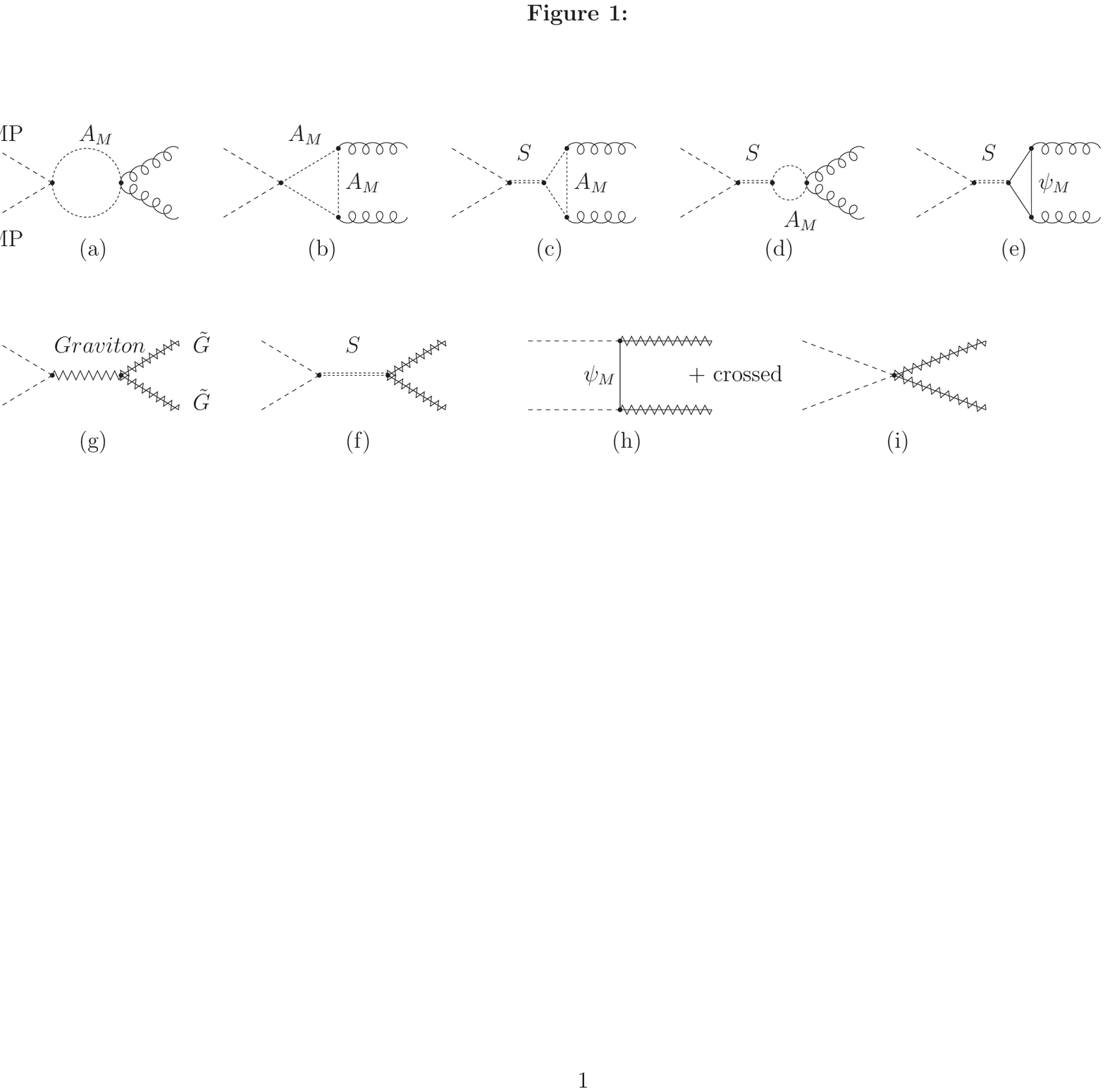}
  \caption{Feynman diagrams of the leading LMP annihilation into $2$
  gluons (a-e) or $2$ gravitinos (f-i).}
\end{figure*}

\begin{figure*}
\label{fig2}
  \includegraphics[height=.3\textheight]{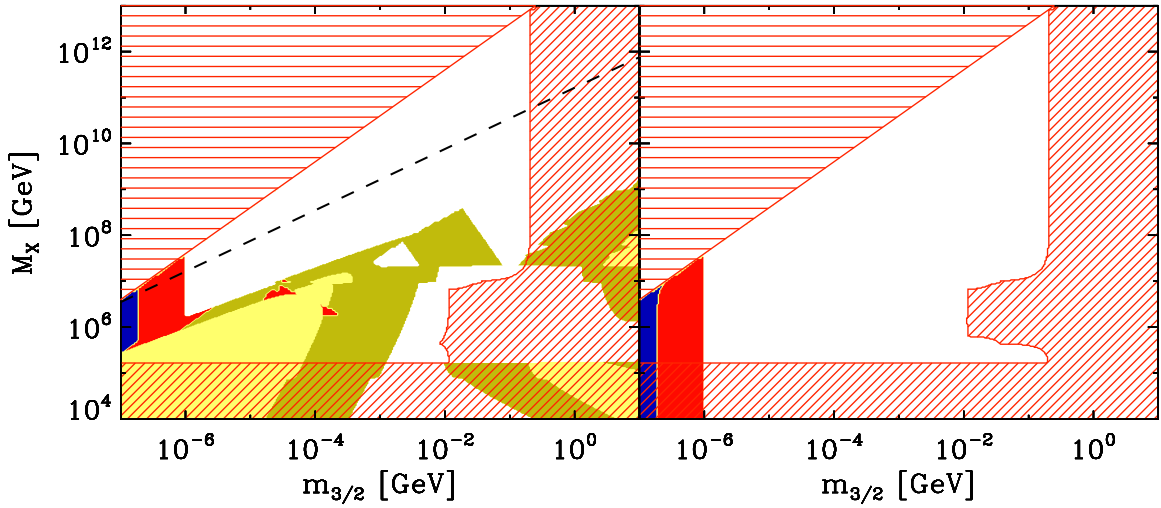}
  \caption{Contours of $\Omega_{3/2}$ in the plane $M_X (\equiv M_-)-m_{3/2}$
  for one pair of messengers sitting in
  $\mathbf{16}+\overline{\mathbf{16}}$ representations of $SO(10)$;
  the LMP is a singlet under $SU(3)\times
  SU(2)\times U(1)$. We take for illustration
  $\kappa^2 \simeq \alpha_s/4\pi$, $f\sim {\cal O}(1)$ and $f' \simeq 5.
  10^{-2}$ and a bino NLSP with $M_{\rm NLSP}=150\,$GeV, $M_{gluino}=1\,$TeV 
  and $k F/M_- \simeq 10^5$GeV;  
  blue (hot), red (warm), green (cold) DM with $0.01 < \Omega_{grav}
  <1$; yellow ($\Omega_{grav} < 0.01$), white ($\Omega_{grav} > 1$). In the 
  right (left) panel the spurion is lighter (heavier) 
  than the messenger. (taken from \cite{JLM04}.)}
\end{figure*}

%

\begin{thebibliography}{999}
%
%
\bibitem{ISS}
 K.~Intriligator, N.~Seiberg and D.~Shih \emph{JHEP 0604:021,2006}.
 
 \bibitem{murayama} see also, H. Murayama these proceedings.
 
\bibitem{GR99} for a review of the earlier GMSB model-building and phenomenology
 see for instance G.F. Giudice, R. Rattazzi, \emph{Phys.Rept 322: 419, 1999},
 and references therein.
  

\bibitem{AK}
S.A.~Abel, C.-S.~Chu, J.~Jaeckel, V.V.~Khoze,  \emph{JHEP 0701:089,2007}, 
\emph{JHEP 0701:015,2007}; N.J.~Craig, P.J.~Fox,  J.G.~Wacker 
\emph{PRD75:085006,2007}; W.~Fischler, V.~Kaplunovsky, C.~Krishnan, 
 L.~Mannelli, M.A.C.~ Torres, \emph{JHEP 0703:107,2007}.
 
\bibitem{DGP}
 S.~Dimopoulos, G.F.~Giudice, A.~Pomarol, \emph{PLB 389:37,1996}. 

\bibitem{ross} L.E.~Ibanez, G.G.~Ross,  
\emph{NPB 368:3-37, 1992.}



\bibitem{JLM04} K. Jedamzik, M. Lemoine, G. Moultaka, \emph{PRD 73:043514,2006};
see also G. Moultaka, \emph{Acta Phys.Polon.B38:645,2007}. 
 
\bibitem{FY02} M. Fujii, T. Yanagida, \emph{PLB 549:273,2002}; see also
E. A. Baltz, H. Murayama, \emph{JHEP 0305:067, 2003}.

\bibitem{JLM05} M. Lemoine, G. Moultaka, K. Jedamzik, 
\emph{PLB 645:222,2007}.

\bibitem{CKLM} M.~Capdequi-Peyran\`ere, M.~Kuroda,  M.~Lemoine, G.~Moultaka,
\emph{to appear}. 

\bibitem{DGP96} S. Dimopoulos, G. F. Giudice, A. Pomarol, \emph{PLB 389:37, 1996}; 
T. Hahn, R. Hempfling, \emph{PLB 415:161, 1997}.

\bibitem{GMSB} M. Dine, A.E. Nelson, \emph{ PRD 48:1277, 1993};
M. Dine, A.E. Nelson, Y. Shirman, \emph{ PRD 51:1362, 1995};
M. Dine {\sl et al.} \emph{PRD 53:2658, 1996}.



\bibitem{SP}
J.~Pradler, F.D.~Steffen, \emph{PLB 648:224,2007}. see also F.D.~Steffen,
these proceedings.

\bibitem{VLHMR} M.~Viel, J.~Lesgourgues, M.G.~Haehnelt, S.~Matarrese, 
A. Riotto, \emph{PRD 71:063534,2005}. 

\bibitem{ETY} M.~Endo, F.~Takahashi, T.T.~Yanagida, \emph{arXiv:hep-ph/0701042},
and \emph{PRD 76:083509,2007} [arXiv:hep-ph/0706.0986v2].

\end{thebibliography}
%

\end{document}